\documentstyle[12pt]{article}
\begin{document}
\begin{center}
{\large\bf An intermediate-mass Higgs boson in two-photon coherent
processes at the LHC}\\
\vspace{.4in}
E. Papageorgiu,\\
LPTHE, Universit\'e de Paris XI, B\^at. 211, F-91405 Orsay.
\end{center}
\vspace{.8in}

\begin{abstract}
\noindent
We reexamine the prospects of searching for a neutral Higgs boson
in the intermediate-mass range, using the proton- and ion-beam
facilities of the LHC to study coherent two-photon processes.
Considering realistic design luminosities
for the different ion beams, we find that beams of
light-to-medium size ions like calcium will give the
highest production rates.
With a suitable trigger and assuming a b-quark identification
efficiency of $30\%$ and a $b{\bar b}$ mass resolution of 10 GeV
one could expect to see a Higgs
signal in the $b{\bar b}$ channel with
a 3-4 $\sigma$ statistical significance in
the first phase of operation of the LHC with beams of calcium.
\end{abstract}

\vskip 2 truecm
\noindent {\bf LPTHE-ORSAY Preprint 95-03}
\vskip 0.5 cm
\noindent
January 1995
\newpage

The search for the Higgs particle of the standard model (SM)
is one of the major scopes of the LHC.
The discovery of such a particle with a mass $M_H$ in the intermediate-mass
range $80 - 180$ GeV, {\it i.e.} beyond the discovery potential of LEP2,
is a challenge for both theory and experiment.
Considerations on the stability of the SM Higgs potential suggest
a strong correlation between the Higgs mass and the energy scale
at which new physics beyond the SM becomes relevant, so $M_H < 100$ GeV
implies a scale of $\sim 10$ TeV [1].
In the case of the minimal supersymmetric standard model (MSSM)
the lightest neutral Higgs scalar mass  should be less than $140$ GeV
for a top-quark mass of $M_t < 190$ GeV or even less than $120$ GeV if
$M_t$ is equal to the infrared fixed-point value [2].
Detailed studies have shown that this lower part of the Higgs mass range
can be explored in pp collisions at the LHC only through a combination
of conventional hard-scattering channels such as
$gg \to H \to \gamma\gamma$, $gg \to WH \to lb{\bar b}$ and $gg \to t{\bar t}H
\to lbb{\bar b}$ and after three years of operation at a luminosity
$L_p \simeq 10^{33} cm^{-2} s^{-1}$ [3].
Even so, to obtain a statistical significance
of more than four one would have to wait until the design luminosity
$L_p^d \simeq 10^{34} cm^{-2} s^{-1}$ is reached by the year $2008$.
Additional Higgs production/decay modes would therefore be welcome.

A different kind of Higgs- and SUSY-particle production,
which would make use of the ion beam facilities at the LHC,
was proposed some time ago [4], namely, the two-photon production in the
coherent electromagnetic field of nucleus nucleus collisions in
which the nuclei $N_{i=1,2}$ would ``ideally'' remain intact:
\begin{equation}
N_1N_2 \to N_1N_2 + X \qquad   X=H,{\bar{\tilde l}}{\tilde l},
{\bar{\tilde\chi}}{\tilde \chi},...\,.
\end{equation}
Triggering on such events
would imply cross sections which scale with the nuclear charges as
$Z_1^2Z_2^2$
and a signal-to-background ratio (S/B) which is sizably more favourable than
in the conventional hard-scattering channels of
\begin{equation}
pp\to X + {\rm jets}\,.
\end{equation}
To illustrate this let us compare the corresponding S/B ratios for
$H\to b{\bar b}$ which would be the predominant SM-Higgs decay mode
if $M_H < 140$ GeV. For the processes of eq.(1) and (2) the
signal and background cross sections factorise into a flux factor
which gives the probability of emitting two photons (gluons) from the nuclei
(protons) and the $\gamma\gamma\to X$ and $gg\to X$ cross sections:
\begin{equation}
\sigma(NN\to NNH) = \sigma(\gamma\gamma \to H \to b{\bar b}) \,\,
M_H^2 {dL_{\gamma\gamma}\over dM_H^2}
\end{equation}
\begin{equation}
\sigma(pp\to H+jets) = \sigma(gg \to H \to b{\bar b}) \,\,
M_H^2 {dL_{gg}\over dM_H^2}
\end{equation}
and
\begin{equation}
{d\sigma \over dW^2}(NN\to NN b{\bar b}) =
\sigma(\gamma\gamma \to b{\bar b}) \,\,
{dL_{\gamma\gamma}\over dW^2}
\end{equation}
\begin{equation}
{d\sigma \over dW^2}(pp\to b{\bar b}+jets) =
\sigma(gg \to b{\bar b}) \,\,
 {dL_{gg}\over dW^2}
\end{equation}
where $W$ is the invariant mass of the $b{\bar b}$ system.
The SM Higgs couples to the gluons mainly through the top-quark loop
$I_t \simeq 1$ for $M_t^2 >> M_H^2$ while there is an extra
contribution from the vector boson loop  $I_W \simeq -1/2$
when it couples to the photons.
Therefore, including the colour factors,
\begin{equation}
\sigma(gg\to H)\simeq {G_F\over 288}\,\,
{\alpha_s^2(M_H^2)\over \sqrt{2} \pi} \,\, |I_t|^2
\end{equation}
over
\begin{equation}
\sigma(\gamma\gamma\to H)\simeq G_F \,\,
{\alpha^2\over \sqrt{2} \pi}\,\, |({2\over 3})^2 I_t + I_W|^2
\end{equation}
is just the ratio of the gauge coupling constants $\alpha_s^2/\alpha^2$,
while
\begin{equation}
\sigma(gg \to b{\bar b}) \simeq {8\pi \alpha_s^2 \over 3 W^2}
\end{equation}
over
\begin{equation}
\sigma(\gamma\gamma \to b{\bar b}) \simeq {4\pi \alpha^2 q^4 \over W^2}
\end{equation}
is enhanced by an extra factor of $10^2$ due to the electric charge of
the b quark.
As a consequence, the signal-to-background ratio in hard-scattering
processes $(S/B)_{HS} \sim \cal{O}(10^{-3}-10^{-4})$ is by 2-3
orders of magnitude less favourable than in coherent processes where
$(S/B)_{COH} \sim \cal{O}(10^{-1})$, assuming a $b{\bar b}$ resolution
of $\cal{R}\simeq 10$ GeV.
These ratios can be improved up to a factor of ten by imposing a
cut on the transverse momentum of the b jets, $p_T \geq 0.4 M_H$,
which affects mainly the softer background spectrum [5].
This brings $(S/B)_{COH}$ down to 1:3 for $120$ GeV $< M_H < 150$ GeV
and to 1:5 for $100$ GeV $< M_H < 120$ GeV [6]. To this, one should add
the large coherent Higgs-boson production which for lead-on-lead collisions
would be of the order of a few tens of picobarns (Fig. 2) and
comparable to the production in hard pp collisions. This is not
surprising because the coherent coupling $Z\alpha$
becomes equal to $\alpha_s$ already for rather light nuclei.

Now, compared to central heavy-ion collisions where one will study the
quark-gluon-plasma signals the occurrence of peripheral collisions
is rather overwhelming. The total cross section  for purely
electromagnetic processes lies in the kilobarn range and it
exceeds the geometric cross section by at least one order of magnitude.
This is an advantage as much as a limitation.
On one hand, in collisions with heavy nuclei there would be
a strong enhancement of the two-photon flux:
\begin{equation}
{dL_{\gamma\gamma}\over dW^2} =  {16\over 3}
{Z_1^2 Z_2^2 \alpha^2 \over\pi^2 W^2}
\times \cal{F}({\gamma \over\sqrt{R_1R_2}W})
\end{equation}
with respect to $e^+e^-$ and $pp$ collisions, in particular,
at the lower end of the mass spectrum $W < \cal{W}_0$
\begin{equation}
\cal{W}_0 =  {2\gamma\over\sqrt{R_1R_2}} \,,
\end{equation}
where
\begin{equation}
\cal{F} \simeq {\rm ln}^3({\cal{W}_0 \over W}) \,.
\end{equation}
At the LHC where the maximum energy of a proton beam will
be $E_p = 7$ TeV, the maximum energy of an ion beam will
be $E_{ion}=E_p Z$ and
\begin{equation}
\gamma \simeq \, 7.5 \, {Z\over A} \,\,[{{\rm TeV \over n}}] \,,
\end{equation}
where $A$ is the atomic number.
This implies that the mass range that could be well explored,
in for example Pb-Pb, Ca-Ca and p-p coherent collisions is
$M_X \leq \cal{W}_0 \simeq$ 168 GeV, 366 GeV and 3 TeV respectively.
The production of masses much higher than $\cal{W}_0$
is exponentially suppressed.
Therefore using heavier ions is not necessarily more advantageous
for exploring the upper part of the mass spectrum,
as the function $\cal{F}$
tends to compensate, partially or fully, the gain from
a higher nuclear charge due to the bigger size $R \simeq r_0 A^{1/3}$
($r_0 \simeq 1.2$ fm) of heavier nuclei.
On the other hand, the maximal achievable  beam luminosity $L_b$ is
limited by intra beam effects which become particularly large
for heavy-ion beams.
The overall advantage of doing two-photon physics with ion beams depends
mainly on $L_b \times dL_{\gamma\gamma}/ dW$ and the efficiency to trigger
on such events.

When this mechanism was first proposed and studied in the context
of an intermediate mass Higgs search it was not known what would
be the maximal achievable luminosity at the LHC for different ion
species, so it was assumed that lead-on-lead collisions would be
the best environment for such searches.
In a recent study [7] it was found that while for a beam
of lead the luminosity would be only
$L_{Pb} \simeq 5\times 10^{26} cm^{-2} s^{-1}$,
for medium-size nuclei like calcium it could be four orders of
magnitude higher $L_{Ca} \simeq 5\times 10^{30} cm^{-2} s^{-1}$.
This gain would not only compensate the loss in the two-photon flux.
As will be shown next, it may allow for an
intermediate-mass Higgs search which looks impossible with lead beams.

Before presenting the results I would like to remark that
knowing precisely the two-photon flux in such collisions
requires a definite prescription for implementing the requirement
of ``coherency'' in the calculations as much as in the experimental set-up.
So it was found that, for the mass range in question,
calculations based on a form factor approach [5]
gave results which were by a factor of $2-10$ higher than
impact-parameter calculations based on a ``hard-disc'' scattering
approach [8-10], the higher masses being more affected by this
uncertainty [8].
The presence of the nuclear elastic form factor is namely not sufficient
to exclude strong interaction processes taking place
after the photons have been emitted from the nuclei.
When nuclei and nucleons overlap, elastic scattering is
partly the ``shadow'' of inelastic processes.
These can be eliminated by imposing cuts on the impact
parameters of the two nuclei:
\begin{equation}
b_i > R_i \qquad {\rm and} \qquad |b^{\to}_1 - b^{\to}_2| > R_1 + R_2 \,,
\end{equation}
{\it i.e}, treating the nuclei as two nonoverlapping, ``hard'' and ``opaque''
discs. The results presented in this paper have been obtained by
using this last approach, - the details of which can be found in
ref.[8] - and are therefore predictions for experiments with an ion
(proton) identification device.

Since the coherency condition of eq.(15) implies transverse momenta
of the final state nuclei
\begin{equation}
q_{iT} < {1\over R_i} \qquad {\rm and} \qquad
(q_1 - q_2)^2_T < {1\over (R_1 + R_2)^2}
\end{equation}
which are limited to a few tens of MeV for heavy nuclei up to a GeV
for protons, tagging may be possible only for the latter.
Alternatively, one will have to veto spectator jets, coming
from diffractive dissociation and/or nucleon fragmentation,
in the very forward and backward direction and with a
transverse momentum of typically a few GeV.
Because in this case one cannot exclude partial or full dissociation of
the nuclei, and, electromagnetic and nuclear interactions taking place
in the final state, the limits in eq.(15) should be relaxed
accordingly. One would then gain back the factor of 2-10 for
the two-photon flux, without getting significant contributions to the
$b{\bar b}$ background, as most such inelastic processes will either
take place below the $b{\bar b}$ threshold or will stem from incoherent
electromagnetic processes.

In Fig. 1 the two-photon luminosity function
$dL/dW = L_b \times dL_{\gamma\gamma}/ dW$
is plotted for collisions of heavy ions (Pb-Pb),
of medium-size ions (Ca-Ca) and protons (p-p), using the values
of ref.[7] for the ion luminosities and $L_p \simeq 10^{33} cm^{-2}s^{-1}$
for the protons. The upper and lower dotted curves correspond to a
nuclear cut-off in eq.(15) of $R \simeq 0.2$ fm, which is the proton
radius as determined from elastic electron scattering,
and $R \simeq 1$ fm respectively. This is to demonstrate how the
absence of a sharp edge for protons and light nuclei (the form factor
is an exponential) may affect the two-photon luminosity function
by choosing a cut-off at the tail of the distribution to ensure
the absence of all strong interaction processes.
For invariant masses up to $250$ GeV coherent collisions with calcium beams
will provide a higher or at least a comparable flux of photons
with respect to proton beams (until the
upgraded luminosity for protons is reached) while the heavy-ion flux
will be two to three orders of magnitude lower.

The total Higgs production cross section for these three type of
collisions is shown in Figs. 2-3. The gain of only one order of magnitude
-on average- in $\sigma(PbPb \to PbPb H)$ over $\sigma(CaCa \to CaCa H)$
multiplied by four orders of magnitude of luminosity loss
means a factor of thousand gain in the event rate when going
from lead to calcium beams.
For $10^7$ sec of running time per year one should expect
$20 - 50$ events per year in Ca-Ca collisions for a Higgs with a mass
of $80 - 180$ GeV. This rate should be compared to the $30-70$ events
from Fig. 3 for coherent p-p collisions after one year of running at the
upgraded luminosity of $L_p^d \simeq 10^{34} cm^{-2} s^{-1}$.

The requirements for detecting a Higgs signal in the $b{\bar b}$ channel
have been studied in detail in the refs.[5,6,11]. Assuming an efficient
trigger for selecting only coherent processes and perfect b identification,
the signal-to-background ratio $(S/B)_{COH}$ is indeed very favourable,
as discussed previously. One should then
expect to see a signal with a  statistical
significance of $S/\sqrt{S+B} \simeq 2$
in the mass range $M_H\simeq 100-130$ GeV
already after one year of running with calcium. On the other hand,
assuming realistic values for the b-detection efficiency
$\epsilon = 30\%$ and for the misidentification
probabilities for a $c{\bar c}$, a $u{\bar u}$, a $d{\bar d}$ and a
$s{\bar s}$ pair of $5\%$, $0.5\%$, $0.5\%$ and $0.5\%$ respectively,
reduces $(S/B)_{COH}$ to $1:5$ in the mass range
$120$ GeV $< M_H < 150$ GeV and to $1:10$ in the mass range
$100$ GeV $< M_H < 120$ GeV [11]. In this case
three to four years of running will be needed to reach a statistical
significance of $3-4$. Even so, this is competitive with
what one does also expect from the hard-scattering channels [3].

These results require however a very efficient trigger. Without it the
background from hard-scattering processes would be
$A^2$ times the pp background of eq.(6). One can estimate, that
\begin{equation}
{\sigma(NN\to NNH) \over A^2 \int_{M_{H}-\cal{R}}^{M_{H}+\cal{R}}
dW^2 d\sigma / dW^2 (pp\to b{\bar b}+jets)} \simeq {(Z\alpha)^4
\over A^2 \alpha_s^4} \, 10^{-3} \, \cal{F}({\cal{W}_0\over M_H})
\end{equation}
is of order $\cal{O}(10^{-6}-10^{-7})$ [11]. Coloured objects like the gluons
don't couple to nuclei or nucleons as such. These processes will
always involve dissociation of the nuclei and partial fragmentation of the
nucleons, leading in the worst of cases to a final state with all
nucleons but one going down the beam pipe and a spectator
jet with $p_T \sim 2-5$ GeV. The rejection of low-$p_T$ jets is
therefore vital.

The background from diffractive processes, with or without dissociation
of the participating nucleons, has not been yet investigated in detail.
Even in deep inelastic scattering at HERA, one percent of all events with a
large rapidity gap are of diffractive origin [12].
The pomeron $\cal{P}$, a colour-singlet object,
is strongly coupled to nucleons
and eventually to nuclei, so that particle production via
double-pomeron and photon-pomeron interactions could a priori
compete with or even exceed the two-photon production in ion-ion
or p-p collisions. Depending on the pomeron structure function,
the $\cal{P}\cal{P}\to H$ exclusive/inclusive cross section
in p-p collisions at the LHC could be as high as $0.1$ pb [13], thus
exceeding by three orders of magnitude the two-photon cross section.
This looks at first very tempting, but, since the irreducible
signal-to-background ratio for $\cal{P}\cal{P}$ processes
is not expected to be any better than for $gg$ processes (the pomeron
is a multigluon state), a full analysis is needed, also, because
the background from these processes could completely swamp the
two-photon signal in p-p collisions [14].
Fortunately, due to nuclear absorption effects, these processes
are strongly suppressed for quasi-elastic collisions of nuclei [15].
In fact, it seems that the A dependence is completely eradicated by this
effect, so that $\cal{P}\cal{P}\to b{\bar b}$ could not obscure the
signal from the electromagnetically produced Higgs bosons.

Taking all this into account it seems that the best place to look
for an intermediate-mass Higgs and study new phenomena in genuin
coherent processes is in collisions of not too heavy nuclei.

\vskip 2mm
\noindent
{\large\bf Acknowledgements}
\noindent
This work was supported by EEC Project: ERBCHBICT930850.

\newpage

\noindent
{\large\bf Figure Captions}

\vskip 2mm
\noindent
{\bf Fig. 1} The two-photon luminosity
$dL/dW = L_b \times dL_{\gamma\gamma}/ dW$
is plotted as a function of the invariant mass $W$
for collisions of lead (Pb-Pb),
calcium (Ca-Ca) and protons (p-p), with
$L_b = 5 \times 10^{26} cm^{-2}s^{-1}$ for $Pb$,
$L_b = 5 \times 10^{30} cm^{-2}s^{-1}$ for $Ca$, as in ref.[7],
and $L_b = 10^{33} cm^{-2}s^{-1}$
for the protons. The upper and lower dotted curves correspond to a
nuclear cut-off in eq.(15) of $R \simeq 0.2$ fm
and $R \simeq 1$ fm respectively.

\vskip 2mm
\noindent
{\bf Fig. 2} The total cross section for the production of the
Higgs particle of the Standard Model through two-photon fusion
in coherent Pb-Pb and Ca-Ca collisions at the LHC.

\vskip 2mm
\noindent
{\bf Fig. 3}  The total cross section for the production of the
Higgs particle of the Standard Model through two-photon fusion
in coherent p-p collisions at the LHC.

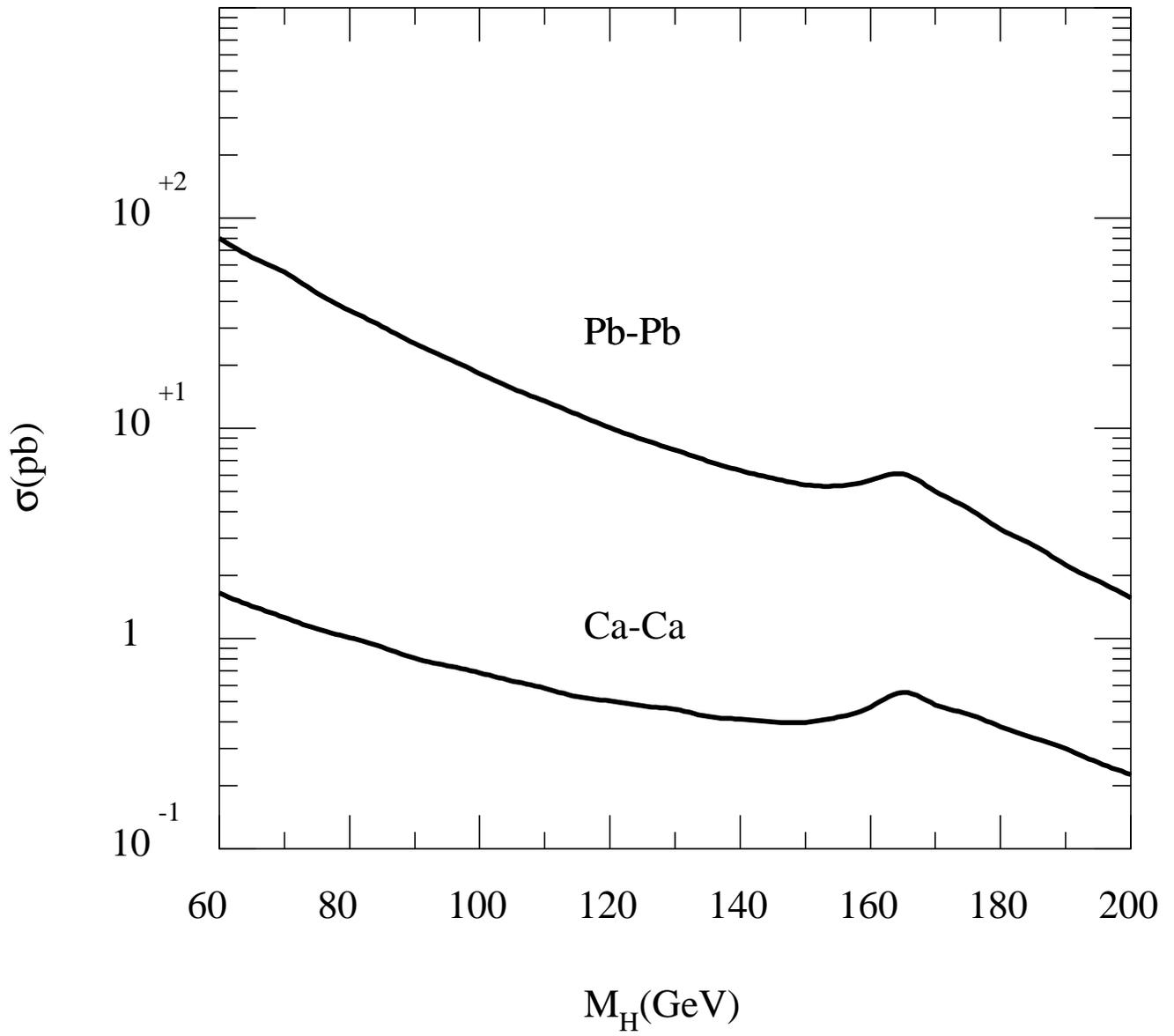
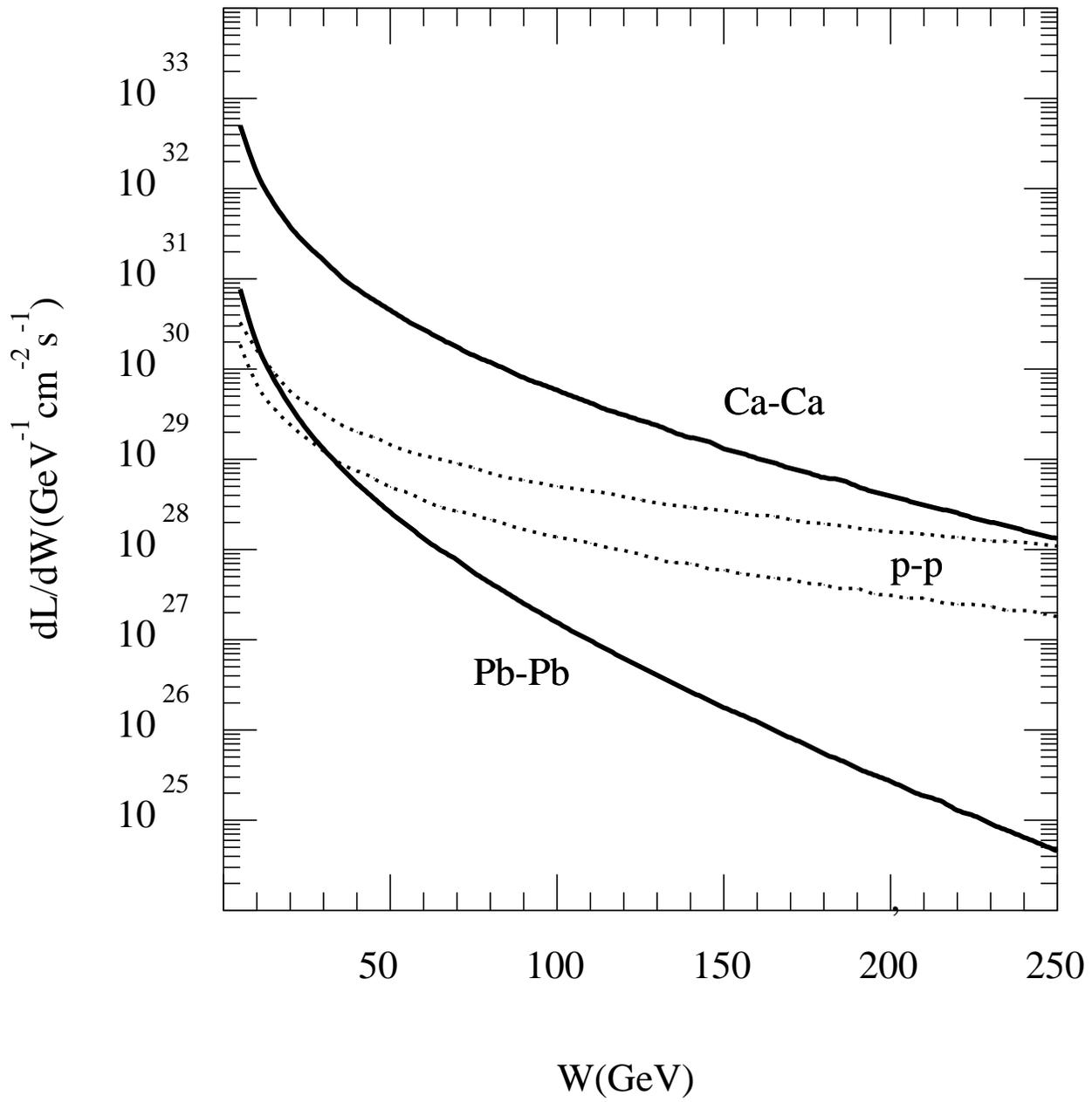
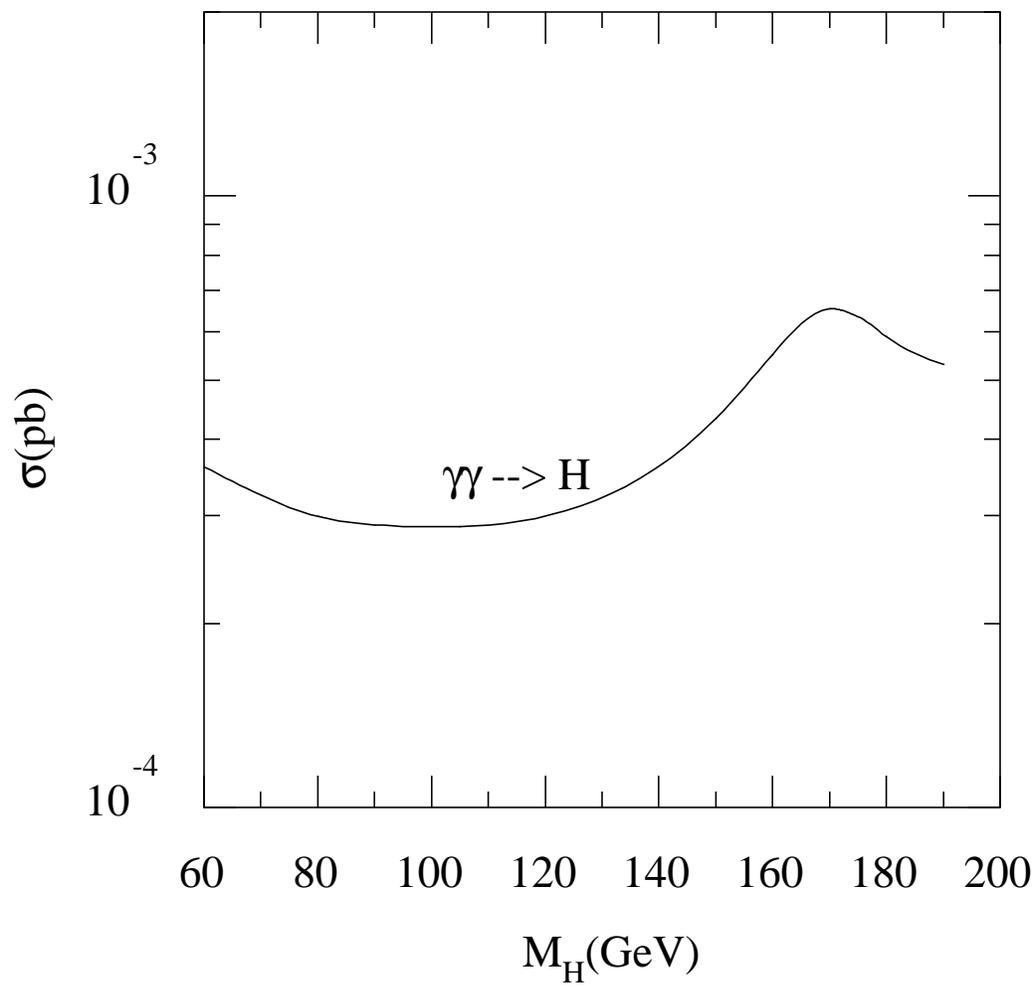
\end{document}